\documentclass[aps,prl,twocolumn,groupedaddress,floats,showpacs]{revtex4-1}
\usepackage{latexsym}
\usepackage{dcolumn}
\usepackage[dvips]{graphicx}
\usepackage{amssymb}
\usepackage{graphics}
\usepackage{amsmath}
\usepackage{epsf}

\newcommand{\beq}    {\begin{equation}}
\newcommand{\enq}    {\end{equation}}

\begin{document}
\title{Coulomb drag in graphene}
\author{E. H. Hwang and S. Das Sarma} 
\affiliation{Condensed Matter Theory Center, Department of 
        Physics, University of Maryland, College Park, MD 20742-4111}

\begin{abstract}
We calculate theoretically the Coulomb drag resistivity for two
graphene monolayers spatially separated by a distance ``$d$''. We show
that the frictional drag induced by inter-layer electron-electron
interaction goes asymptotically as $T^2/n^3$ and $T^2 \ln(n)/n$ in the
high-density ($k_F d \gg 1$) and low-density ($k_F d \ll 1$) limits,
respectively. 
\end{abstract}
\maketitle


Frictional drag measurements of transresistivity in
double layer systems have led to significant advances in our
understanding of density and temperature dependence of
electron-electron interactions in 2D systems \cite{hwang:2003}.
Recent interest has focused on the role of
electron interaction effects on the graphene drag resistivity, which should vary in a
systematic manner as a function of electron density ($n$), layer separation
($d$),  and temperature ($T$).
In particular, a recent experiment of Coulomb drag in double layer
graphene by Kim {\it et al}. \cite{kim:2011} 
is particularly interesting.
In view of the considerable fundamental significance of the issues
raised by the experimental observations, we present in this paper a
careful theoretical calculation of frictional drag $\rho_D(T)$ in a
2D \cite{hwang:2009} graphene within the canonical many-body Fermi
liquid theory. The current work is a generalization of the earlier
theoretical work on graphene drag by Tse {\it et al.} \cite{tse:2007}.


We start by writing down the theoretical formula for
$\rho_D$ \cite{tse:2007} in the
many-body Fermi liquid 
RPA-Boltzmann theory approximation widely used in the literature.
The double layer frictional drag in a many body-Fermi liquid  
diagrammatic perturbation theory with dynamically screened
electron-electron interaction is given by \cite{tse:2007}
\begin{equation}
\rho_D=\frac{\hbar^2}{2 \pi e^2n^2 k_BT}\int\frac{q^2 d^2q}{(2\pi)^2}
\int{d\omega} \frac{F_1(q,\omega)F_2(q,\omega)}
{\sinh^2(\beta \omega/2)},
\label{eq:rhod}
\end{equation}
where $F_{1,2}(q,\omega) = |u_{12}^{sc}(q,\omega)| \rm{Im}
\Gamma_{11,22}(q,\omega)$, with $u_{12}^{sc} = v_{12}^{c}
/\epsilon(q,\omega)$ is the dynamically screened interlayer Coulomb
interaction between layers, and $\Gamma(q,\omega)$ is the 2D graphene
non-linear susceptibility\cite{tse:2007}. The interlayer Coulomb interaction is given by
$v_{12}^{c}(q)= v^{c}(q)e^{-qd}$ with the intralayer Coulomb potential \cite{hwang:2009}
$v_c(q) = 2\pi e^2/\kappa q$ where $\kappa$ is the background
dielectric constant. (We consider the so-called balanced situation here with
the same carrier density $n$ in both layers.) Note that the dielectric
function ${\epsilon}(q,\omega)$ entering
Eq.~(\ref{eq:rhod}) is given by \cite{hwang:2009}
\begin{eqnarray}
|\epsilon(q,\omega)|& = &\left [1-v_{11}(q)\Pi_{11}(q,\omega) \right ]
\left [1-v_{22}(q)\Pi_{22}(q,\omega) \right ] \nonumber \\
& - & v_{12}(q)v_{21}(q)\Pi_{11}(q,\omega)\Pi_{22}(q,\omega),
\label{eq:eps}
\end{eqnarray}
where $v_{ii}(q) = v_c(q)$ and $\Pi_{ii}$ is the intralayer graphene
polarizability \cite{hwang:2007}. 
Even tough there are analytic expressions for graphene polarizability
at $T=0$,\cite{hwang:2007} the finite temperature versions of the polarizability have
not been calculated analytically. The full expression of finite temperature
polaizability is necessary to understand more precisely the 
temperature dependent drag including the plasmon enhancement effects\cite{hu:1995}. Here we
provide the efficient way for calculating the finite 
temperature graphene polarizability by generalizing our earlier work \cite{hwang:2007}
\begin{widetext}
\begin{equation}
\Pi(q,w,T) = \frac{\pi}{8}\frac{q^2}{\sqrt{q^2-\omega^2}} +
\int_{0}^{\infty}dk \left[ f(q) +g(q) \right ] \left [ \frac{ \left ( 
      \frac{\omega^2-q^2}{4}+\omega k + k^2 \right ) {\rm sgn}(a_{+}) }{ 
\sqrt{\frac{(\omega^2-q^2)^2}{4} +  (\omega^2-q^2) (\omega k +
    k^2)}}  +
\frac{\left ( \frac{\omega^2-q^2}{4}- \omega k + k^2 \right )
{\rm  sgn}(a_{-})  }{ 
\sqrt{\frac{(\omega^2-q^2)^2}{4} +  (\omega^2-q^2) (-\omega k +
    k^2)}}  
\right ],
\label{eq:pit}
\end{equation}
\end{widetext}
where $\Pi = \Pi/D_0$ ($D_0=2k_F/\pi \hbar v_F^2$ is the density of
states of graphene at Fermi energy, and $k_F$ and $v_F$
are the Fermi wave vector and Fermi velocity
of graphene), $q=q/k_F$, $\omega=\omega/E_F$,
$f(q) = [e^{-(\varepsilon_q-\mu)}-1]^{-1}$, $g(q) =
[e^{(\varepsilon_q+\mu)}+1]^{-1}$, $\varepsilon_q=\hbar v_F q$, and
$a_{\pm} = \omega^2 - q^2 \pm 2\omega k$. In $f(q)$ and $g(q)$, $\mu$
is the finite temperature chemical potential which must be calculated
self-consistently to
conserve the total electron density.

\begin{figure}
\includegraphics[width=0.8\columnwidth]{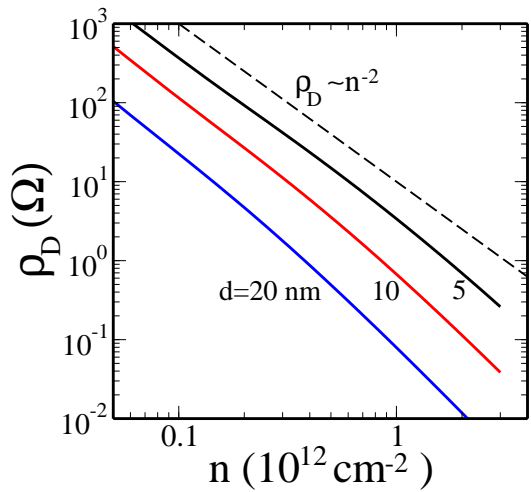}
\caption{
The density dependent Coulomb drag 
for different layer separations $d=5$, 10, 20 nm and $T=200K$.
Dashed line indicates $\rho_D \sim n^{-2}$ behavior. 
\label{fig:2}
}
\end{figure}

With assumptions of a large
inter-layer separation ($k_{F}d\gg 1$, or $q_{TF}d\gg 1$, with $q_{TF}$
being the Thomas Fermi (TF) screening wave 
vector) and the random phase approximation (RPA) in which $\Pi
_{ii}$ is replaced by its value for the non-interacting electrons, we have 
for the identical layers at high density and low temperature,
\begin{equation}
\rho_{D}=\frac{h}{e^2}
\frac{\zeta (3)\pi }{16}
  \frac{(T/T_F)^2}{(k_{F}d)^{2}(q_{TF}d)^{2}} 
\label{eq:anal1}
\end{equation}
where $q_{TF}=4 r_s k_F$ 
is the TF wave vector with the graphene fine structure constant
$r_s = e^2/\kappa \hbar v_F$ 
and $\zeta$ is the Riemann zeta function. 
This result shows that $\rho_D(n) \propto n^{-3}$, but this result,
which was obtained in ref. \cite{tse:2007},
applies only for high density and large separation limits (or weak interlayer
correlation, $k_Fd \gg 1$). At low
densities (or strong interlayer correlation, $k_F d \ll 1$) the exponent
in the density dependent drag   
differs from -3 as shown in Fig.~\ref{fig:2} where we directly
numerically calculated graphene drag using Eqs.~(\ref{eq:rhod})-(\ref{eq:pit}).
Eq.~\ref{eq:anal1} shows $\rho_D(T) \propto T^{2}$.
For large layer separation (i.e. $k_Fd \gg 1$)
the back-scattering $q \approx 2k_F$ is suppressed
due to the exponential dependence of the interlayer Coulomb
interaction $v_{12}(q) \propto \exp(-qd)/q$ as well as the graphene
chiral property. In this case the drag is
dominated by small angle scattering and one expects $\rho_D \propto
T^2/(n^3 d^4)$. 

For the strong interlayer correlation ($k_F d \ll1$) in the
low-density or small-separation limit, we have the
following asymptotic behavior:
\begin{equation}
\rho_D =\frac{h}{e^2}\frac{8\pi r_s^2}{3}\frac{T^2}{T_F^2} \ln \left [
  \frac{(2q_{TF}d +1)^2}{4q_{TF}d(1+q_{TF}d)} \right ].
\end{equation}
Thus, we have $\rho_D(T) \sim T^2$ again, and $\rho_D(n) \sim
\ln(n)/n$ with very weak logarithmic $d$-dependence.


\begin{figure}
\includegraphics[width=\columnwidth]{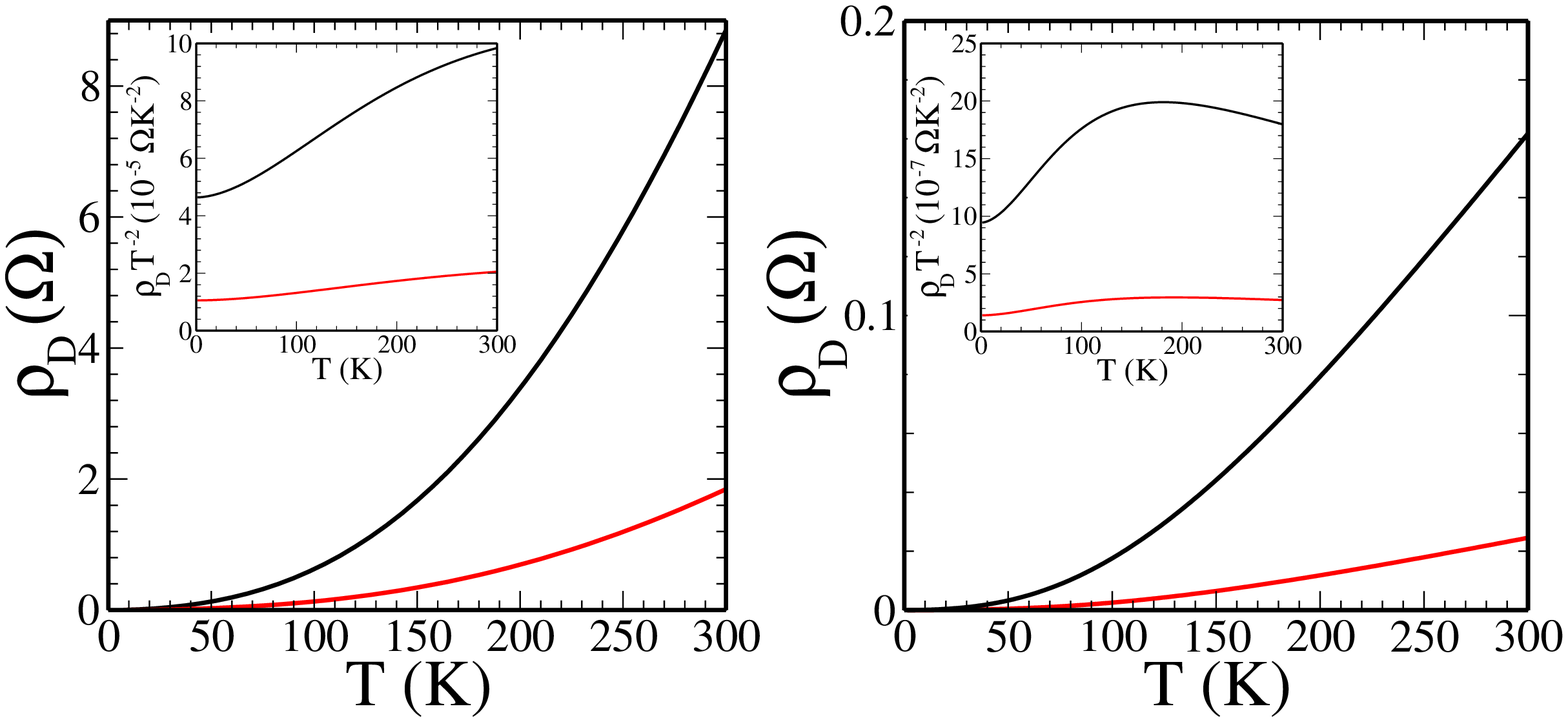}
\caption{
The temperature dependence of Coulomb drag 
for (a) $d=50$\AA \;and (b) $d=200$\AA\;. The black (red) lines are
the results calculated with equal electron densities, $n_1=n_2=10^{12}cm^{-2}$
($n_1=n_2=2\times 10^{12}cm^{-2}$).
Insets show the scaled drag resistivity $\rho_D(T)/T^2$.
\label{fig:1}
}
\end{figure}

In Fig.~\ref{fig:1} we show the calculated Coulomb drag as a function
of temperature for two different densities 
$n=1$, $2\times 10^{12}cm^{-2}$, layer separation (a) $d=50$ {\AA} (b)
$d=200$ {\AA}. The overall temperature dependence of drag is  
close to the quadratic behavior. But we find a small
corrections, especially at low values of $k_Fd$.
In regular 2D systems there is  a $\ln(T)$ corrections to the $T^2$ dependence
of the drag. However, due to the suppression of the back-scattering in
graphene such logarithmic
correction does not show up in our numerical results except perhaps at
extremely low temperatures.

In Fig.~\ref{fig:2} the density dependent Coulomb drag is shown for
different layer separations.
Our calculated Coulomb drag resistivity 
follow a $n^{\alpha}$ dependence with $\alpha \alt -2$ at low carrier
densities (or, $k_F d <1$), 
but as the density increases the exponent ($\alpha$) decrease.
Based on our calculation we believe that the experimental departure from the $n^{-3}$
behavior reported in Ref.~\cite{kim:2011} is essentially a
manifestation of the fact that 
the asymptotic $n^{-3}$ regime is hard to reach in low density electron
systems where $k_F d \gg 1$ limit simply cannot be accessed. We
predict a weak $\ln(n)/n$ density dependence in the low-density or
small separation limit.

In conclusion, we study the frictional drag 
between two spatially separated graphene layers within a many-body Fermi liquid
theory. 
We find that the temperature dependent drag mostly shows a quadratic
behavior regardless of the layer separation, but
the density dependence varies from $\ln(n)/n$ for $k_F d \ll 1$ to
$n^{-3}$ for $k_F d \gg 1$. But most currently available double layer
graphene samples belong to
$k_F d \sim 1$, so the density dependence dose not have any universal
power law behavior.
We also find that due to the suppression of the $q=2k_F$
back-scattering there is no $\ln(T)$ correction in the drag
resistivity in graphene.

This work is supported by the US-ONR.


%

\end{document}